\documentclass[reprint,amsmath,amssymb,aps]{revtex4-2}

\usepackage{graphicx}% Include figure files
\usepackage{amsmath,mathtools}
\usepackage{comment}
\usepackage{gensymb}
\usepackage{dcolumn}
\usepackage{tikz}
\usepackage{pgfplots}
\usepackage{catchfilebetweentags}
\usepackage{gensymb}
\usepackage{float}

\newcolumntype{d}[1]{D{.}{.}{#1}}

\usepackage[version=4]{mhchem}

\begin{document}

\title{Enhancing Density Functional Theory for Static Correlation in Large Molecules}

\author{Daniel Gibney}
\author{Jan-Niklas Boyn}%
\author{David A. Mazziotti}%

\email{damazz@uchicago.edu}%

\affiliation{Department of Chemistry and The James Franck Institute, The University of Chicago, Chicago, Illinois 60637 USA}%

\date{Submitted July 13, 2024\textcolor{black}{; Revised September 16, 2024}}

\begin{abstract}

A critical challenge for density functional theory (DFT) \textcolor{black}{in practice} is its \textcolor{black}{limited} ability to treat static electron correlation, leading to errors in its prediction of charges, multiradicals, and reaction barriers.  Recently, we combined one- and two-electron reduced density matrix theories with DFT to obtain a universal $O(N^3)$ generalization of DFT for static correlation.  In this Letter, we enhance the theory's treatment of large molecules by renormalizing the trace of the two-electron identity matrix in the correction using Cauchy-Schwarz inequalities of the electron-electron repulsion matrix. We apply the resulting functional theory to linear hydrogen chains as well as the prediction of the singlet-triplet gap and equilibrium geometries of a series of acenes. This renormalization of the generalized DFT retains the $O(N^{3})$ computational scaling of DFT while enabling the accurate treatment of static correlation for a broad range of molecules and materials.

\end{abstract}

\maketitle

\textit{Introduction:} Despite its ability to treat the electron correlation of many molecular systems with mean-field computational cost, density functional theory (DFT)~\cite{Parr.1994, Dreizler.1990, Jones.2015, Hohenberg.1964} has limitations in its treatment of charges~\cite{Dreuw.2004, Steinmann.2012}, barrier heights~\cite{Kaplan.2023}, and bi- and multiradicals~\cite{Cohen.2008}. These limitations arise from the inability of the approximate functionals employed within DFT to provide a full description of static, or multireference, electron correlation. Static correlation occurs in molecular systems with energetically degenerate, or nearly degenerate, orbitals that lead to two or more Slater determinants contributing substantially to the wave function~\cite{Montgomery.20188q7}.  Recently, we combined DFT~\cite{Parr.1994, Dreizler.1990, Jones.2015, Hohenberg.1964} and its extensions~\cite{Kulik.2006, Chai.2012, Manni.2014, Li.2017, Lee.2019, Gibney.2021, Bajaj.2021, Mei.2021, Su.2021, Gibney.20225zn, Seenithurai.2023} with 1-electron  and 2-electron reduced density matrix (1-RDM~\cite{Gilbert.1974, Levy.1979, Valone.1980, Muller.1984, Goedecker.1998, Mazziotti.2000q02, Piris.2007, Sharma.2008, Rohr.2008, Piris.2017, Schilling.2019, Schmidt.2019, Piris.2021, Schilling.2021, Wang.2022, Gibney.2022, Lemke.2022, Liebert.2023} and 2-RDM~\cite{Coleman.2000, MazziottiB.2007, Colmenero.1993b, Nakatsuji.1996, Mazziotti.1998e39, Mazziotti.1998, Mazziotti.1998bme, Mukherjee.2001, Mazziotti.20060v3, Alcoba.2011, Boyn.2021, Smart.2021, Kutzelnigg.2006, Sokolov.2013, Mazziotti.2008p0g, Sand.2012, Mazziotti.2001g, Nakata.2001, Zhao.2004, Mazziotti.2004, Shenvi.2010, Verstichel.2011, Mazziotti.20115w, Mazziotti.2016co, Li.2021, Knight.2022, Xie.20228s, Mazziotti.2023}) theories to obtain a universal $O(N^3)$ generalization of DFT for static correlation~\cite{Gibney.2023}.  This generalization transforms DFT into a 1-RDM functional theory (RDMFT) whose convexity allows the orbital occupations to become fractional.

In this Letter, we enhance the generalization of DFT for an improved treatment of static correlation in large molecules.  The generalized DFT, or RDMFT, that we previously derived~\cite{Gibney.2023}, involves the trace of the two-electron identity matrix and the trace of the cumulant part of the 2-RDM~\cite{Mazziotti.1998e39, Mazziotti.1998, Mazziotti.1998bme, Mukherjee.2001, Mazziotti.2002cv9, Coleman.1980}.  The former, however, scales quadratically with system size $r$ while the latter scales linearly with $r$.  This scaling mismatch causes the correction to the DFT energy to become less effective for larger molecules.  Here we renormalize the trace of the identity matrix by reweighting the two-electron space with the electron repulsion term $1/r_{12}$, reducing its scaling from quadratic to linear.  The reweighting is performed with Cauchy-Schwarz inequalities~\cite{Whittaker_Watson_1996} that relate the magnitudes of the diagonal and off-diagonal elements of the two-electron repulsion matrix. To demonstrate the effectiveness of this approach, we apply the corrected functional theory to linear hydrogen chains~\cite{Sinitskiy.2010} as well as the prediction of the singlet-triplet gap and equilibrium geometries of acene chains~\cite{Hachmann.2007, Gidofalvi.2008, Hemmatiyan.2019}.  The renormalization results in an RDMFT, solvable by semidefinite programming~\cite{Mazziotti.20115w}, that accurately treats static correlation for a broad range of molecules and materials at the same $O(N^{3})$ computational scaling as DFT.

\textit{Theory:} Following a brief review of the correction in the generalized DFT, we derive the renormalization of the correction for large molecules.  The energy of an $N$-electron atom or molecule in a finite basis of $r$ spin orbitals can be written as a functional of the 1-RDM and the cumulant (or connected) part of the 2-RDM~\cite{Mazziotti.2000q02}
\begin{equation}
\label{eq:rdmft}
E_{\rm RDMFT}[^{1} D, ^{2} \Delta] = E[^{1} D] + E^{\Delta}[^{2} \Delta]
\end{equation}
in which
\begin{eqnarray}
\label{eq:e1}
E[^{1} D] & = & {\rm Tr}(^{1} H \, ^{1} D) + {\rm Tr}(^{2} V \, ({^{1} D \wedge {^{1}D}})) \\
E^{\Delta}[^{2} \Delta] & = & {\rm Tr}(^{2} V \, ^{2} \Delta) \label{eq:cumulante}
\end{eqnarray}
and
\begin{equation}
\, ^{2} \Delta = \, ^{2} D - \, ^{1} D \wedge {^{1}D} \,
\end{equation}
where $^{1} H$ is the matrix representation of the one-electron kinetic energy and nuclear-electron Coulomb terms, $^{2} V$ is the matrix representation of the two-electron repulsion term, $^{1} D$ and $^{2} D$ are the 1- and 2-RDMs, normalized to $N$ and $N(N-1)/2$, respectively, $\wedge$ denotes the antisymmetric (or Grassmann) tensor product~\cite{Mazziotti.1998e39} and $^{2} \Delta$ is the cumulant part of the 2-RDM~\cite{Mazziotti.1998e39, Mazziotti.1998, Mazziotti.1998bme, Mukherjee.2001}.

The cumulant 2-RDM can be decomposed into three orthogonal subspaces based on the unitary group~\cite{Mazziotti.2002cv9}, known as the unitary decomposition~\cite{Coleman.1980}
\begin{equation}
^{2} \Delta = {^{2} \Delta_{0}} + {^{2} \Delta_{1}} + {^{2} \Delta_{2}} ,
\end{equation}
\color{black}
which upon substitution into Eq.~(\ref{eq:cumulante}) yields a decomposition of the cumulant energy into zero-, one-, and two-body components
\begin{equation}
E^{\Delta}[^{2} \Delta] = E^{\Delta}[^{2} \Delta_{0}] + E^{\Delta}[^{2} \Delta_{1}] + E^{\Delta}[^{2} \Delta_{2}] .
\end{equation}
 \color{black}
By approximating the cumulant part of the energy in terms of just the zeroth component of the unitary decomposition, we have
\begin{equation}
E^{\Delta}[^{2} \Delta] \approx  \color{black} E^{\Delta}[^{2} \Delta_{0}] =  \color{black} {\rm Tr}(^{2} V \, {^{2} \Delta_{0}}).
\end{equation}
Further substituting the explicit unitary decomposition~\cite{Mazziotti.2002cv9} in which $^{2} I$ is the two-electron identity matrix
\begin{equation}
^{2} \Delta_{0} = \frac{1}{{\rm Tr}(^{2} I)} {\rm Tr}(^{2} \Delta) \, ^{2} I
\end{equation}
yields
\begin{equation}
\label{eq:E0}
E^{\Delta}[^{2} \Delta] \approx \frac{1}{{\rm Tr}(^{2} I)} {\rm Tr}(^{2} V) {\rm Tr}(^{2} \Delta) .
\end{equation}
Finally, the trace of the cumulant 2-RDM can be expressed in terms of the 1-RDM's idempotency~\cite{Sajjan.2018yal, Raeber.2015, Mazziotti.2000yx, Valdemoro.1992}
and the trace of ${^{2} V}$ can be expressed in terms of the two-electron repulsion integrals in physicist's notation to obtain
\begin{equation}
\label{eq:ce}
E^{\Delta}[^{2} \Delta] \approx - \gamma {\rm Tr}( {^{1} D} - {^{1} D}^{2})
\end{equation}
where
\begin{equation}
\label{eq:cg}
\gamma = \frac{1}{{\rm Tr}(^{2} I)} \sum_{{\tilde i},{\tilde j}}{ \left ( 2 \langle {\tilde i} {\tilde j} || {\tilde i} {\tilde j} \rangle -  \langle {\tilde i} {\tilde j} || {\tilde j} {\tilde i} \rangle \right ) }
\end{equation}
in which the tilde denotes the index of the spatial part of the spin orbital.  Equation~(\ref{eq:ce}), derived in Ref.~\cite{Gibney.2023}, provides a 1-RDM approximation for the cumulant energy that in combination with the effective one-electron energy $E[^{1} D]$ in Eq.~(\ref{eq:rdmft}) yields a 1-RDM functional theory that corrects the Hartree-Fock energy for static correlation.

As shown in previous work~\cite{Gibney.2023}, we can apply this correction to transform DFT rather than Hartree-Fock into an RDMFT.  To treat DFT with Eq.~\ref{eq:rdmft}, we first modify the one-body energy in Eq.~(\ref{eq:e1}) to include the exchange-correlation functional $F_{\rm xc}[\rho]$
\begin{equation}
E[^{1} D] = T[^{1} D] + V[\rho] + F_{\rm xc}[\rho]
\end{equation}
where $\rho$ is the one-electron density, $T[^{1} D]$ is the interacting kinetic energy functional, and $V[\rho]$ is the external potential including the electron-nuclei Coulomb potential and the Hartree-Fock Coulomb potential.  Second, because an approximate exchange-correlation functional already includes some \textcolor{black}{static} correlation, we define the cumulant energy in Eq.~(\ref{eq:ce}) with the same functional form but a different weight parameter $w$
\begin{equation}
\label{eq:cew}
E^{\Delta}[^{2} \Delta] \approx - w {\rm Tr}( {^{1} D} - {^{1} D}^{2})
\end{equation}
where $w = \kappa \gamma$ in which $\gamma$ is defined as before in Eq.~(\ref{eq:cg}) and $\kappa$ is a damping factor such as $\kappa \in [0,1]$.  When $\kappa=1$, the cumulant energy in Eq.~(\ref{eq:cew}) is identical to the cumulant energy in Eq.~(\ref{eq:ce}), but when $\kappa<1$, the cumulant energy in Eq.~(\ref{eq:cew}) provides a smaller correction than that in Eq.~(\ref{eq:ce}), reflecting that an approximate exchange-correlation functional contains some static correlation.  Importantly, as observed previously in Ref.~\cite{Gibney.2023}, the $\kappa$ is largely independent of the molecular system because the system-dependent behavior is captured by $\gamma$, and thus, for a given approximate density functional a single value for $\kappa$ can be used across molecules.  \textcolor{black}{The $\kappa$ is also similarly independent of both molecular geometry including dissociation and molecular size because the change in static correlation with geometry and size is included in the correction's explicit dependence on the 1-RDM and $\gamma$.}  We find that the optimal $\kappa$ for a given functional increases linearly with the amount of Hartree-Fock exchange in the functional.  The advantage of using the cumulant energy as a correction to DFT rather than Hartree-Fock theory is that DFT captures the dynamic correlation.

For large molecules, however, the cumulant energy in Eq.~(\ref{eq:ce}) generally does not recover enough of the static correlation.  The problem arises because the trace of the two-electron identity matrix in $\gamma$,
\begin{equation}
{\rm Tr}(^{2} I) = \frac{r(r-1)}{2},
\end{equation}
scales quadratically, rather than linearly, in $r$.  Each of the diagonal elements of the identity matrix $^{2} I^{ij}_{ij}$ is equal to one even if the spin orbitals $i$ and $j$ correspond to spatial orbitals that are significantly separated in space.  In taking the trace of the identity matrix, we need to reweight the two-electron space to account for the locality of the electron repulsion matrix due to the locality of the Coulomb repulsion.

To perform the reweighting, we consider the electron repulsion matrix in a local orbital basis, such as the atomic orbitals, with the indices arranged in chemist's notation, denoted by $^{2} {\tilde V}$
\begin{equation}
^{2} {\tilde V}^{{\tilde i}{\tilde i}}_{{\tilde j}{\tilde j}} = \int{ \frac{|\chi_{{\tilde i}}(1)|^{2} |\chi_{{\tilde j}}(2)|^{2}}{r_{12}} d1 d2 }  .
\end{equation}
Because this matrix $^{2} {\tilde V}$ is positive semidefinite, that is $^{2} {\tilde V} \succeq 0$, its diagonal and off-diagonal elements must obey the Cauchy-Schwarz inequalities~\cite{Whittaker_Watson_1996}
\begin{equation}
\left | ^{2} {\tilde V}^{{\tilde i}{\tilde i}}_{{\tilde j}{\tilde j}} \right |^{2} \le
{^{2} {\tilde V}}^{{\tilde i}{\tilde i}}_{{\tilde i}{\tilde i}} \, {^{2} {\tilde V}}^{{\tilde j}{\tilde j}}_{{\tilde j}{\tilde j}} .
\end{equation}
Dividing the left-hand side by the right-hand side of the inequalities and taking the square root, we obtain  weights ${\tilde W}_{{\tilde i}{\tilde j}}$ of the two-electron space
\begin{equation}
{\tilde W}_{{\tilde i}{\tilde j}} = \frac{^{2} {\tilde V}^{{\tilde i}{\tilde i}}_{{\tilde j}{\tilde j}}}{\sqrt{^{2} {\tilde V}^{{\tilde i}{\tilde i}}_{{\tilde i}{\tilde i}} \, ^{2} {\tilde V}^{{\tilde j}{\tilde j}}_{{\tilde j}{\tilde j}}}} = \frac{\langle {\tilde i} {\tilde j} || {\tilde i} {\tilde j} \rangle}{\sqrt{\langle {\tilde i} {\tilde i} || {\tilde i} {\tilde i} \rangle \langle {\tilde j} {\tilde j} || {\tilde j} {\tilde j} \rangle}}
\end{equation}
which lie between 0 and 1
\begin{equation}
0 \le {\tilde W}_{{\tilde i}{\tilde j}} \le 1 .
\end{equation}
When ${\tilde i}={\tilde j}$, the weights equal one (${\tilde W}_{{\tilde i}{\tilde i}} = 1$), but when the local spatial orbitals, ${\tilde i}$ and ${\tilde j}$, are far from each other, the weights are much less than one (${\tilde W}_{{\tilde i}{\tilde j}}\ll1$). The nonzero elements of the weight matrix in the local spin orbital basis set $^{2} W$ can be defined as
\begin{equation}
^{2} W^{ij}_{ij} = {\tilde W}_{{\tilde i}{\tilde j}}
\end{equation}
with ${\tilde i}$ denoting the spatial orbital associated with the spin orbital of $i$.  Using these weights on the diagonal elements of the two-electron identity matrix renormalizes its trace to yield
\begin{equation}
{\rm Tr}(^{2} W \, ^{2} I) = 4 \sum_{{\tilde i}<{\tilde j}}{{\tilde W}_{{\tilde i}{\tilde j}}} + \sum_{{\tilde i}}{{\tilde W}_{{\tilde i}{\tilde i}}} .
\end{equation}
Asymptotically, the trace of the renormalized two-electron identity matrix scales linearly with system size, and therefore, substituting this identity matrix for the conventional identity matrix in Eq.~(\ref{eq:cg}) yields an approximation for the cumulant energy, similar in form to  Eq.~(\ref{eq:ce}), that scales linearly with system size
\color{black}
\begin{equation}
\label{eq:ce3}
E^{\Delta}[^{2} \Delta] \approx - {\tilde \gamma} {\rm Tr}( {^{1} D} - {^{1} D}^{2})
\end{equation}
where
\begin{equation}
\label{eq:cg3}
{\tilde \gamma} = \frac{1}{{\rm Tr}(^{2} W \, ^{2} I)} \sum_{{\tilde i},{\tilde j}}{ \left ( 2 \langle {\tilde i} {\tilde j} || {\tilde i} {\tilde j} \rangle -  \langle {\tilde i} {\tilde j} || {\tilde j} {\tilde i} \rangle \right ) }
\end{equation}
in which the \color{black} ${\tilde \gamma}$ is related to the original $\gamma$ by a ratio of the traces of the two identity matrices
\begin{equation}
\label{eq:cg2}
{\tilde \gamma} = \frac{{\rm Tr}(^{2} I)}{{\rm Tr}(^{2} W \, ^{2} I)} \gamma .
\end{equation}
Similarly, we define the renormalized weight ${\tilde w}$ as ${\tilde w} = {\tilde \kappa} {\tilde \gamma}$ in which ${\tilde \kappa}$ is a modified damping parameter.  Using these modified parameters, we can obtain accurate corrections for static correlation with either Hartree-Fock theory or DFT for both large and small molecules.

\textcolor{black}{The renormalization makes the energies of large molecules, exhibiting appreciable decay in their electron-electron repulsion, consistent with the energies of smaller molecules, not exhibiting such decay.  By correcting the energy scaling with molecular size, the reweighting---or renormalization---also corrects the convexity of the functional.  For example, correcting the energy of two molecules at infinite separation can be viewed as a correction to the convexity of the functional with respect to the RDMs of the individual molecules.  We further note that the cumulant energy in Eq.~(\ref{eq:cumulante}), without approximation, exhibits the correct linear scaling with molecular size and hence, the renormalization restores this correct scaling to the approximated cumulant energy.}

\textit{Results:} We apply the RDMFT with both the original $w$ and renormalized ${\tilde w}$ weights to the stretching of linear hydrogen chains and the singlet-triplet gap as well as the carbon-carbon bond lengths of acene chains ranging in length from 5 to 12 units. We use $w = \kappa\gamma$ and $\Tilde{w} = \Tilde{\kappa} \Tilde{\gamma}$ with $\gamma$ and $\Tilde{\gamma}$ being obtained from Eq.~(\ref{eq:cg}) and Eq.~(\ref{eq:cg2}), respectively. All RDMFT calculations are performed with the correlation-consistent polarized valence double-zeta (cc-pVDZ) basis set ~\cite{Dunning.1989} and the SCAN functional~\cite{Sun.2015} with $\kappa = 0.158$, as employed in our previous work~\cite{Gibney.2023}, and $\Tilde{\kappa} = 0.112$.  Analytical gradients~\cite{Weinert1992} are implemented for geometry optimizations.  The RDMFT calculations are solved using a self-consistent-field (SCF) procedure at $\mathcal{O}(N^3)$ scaling, described in Refs.~\cite{Gibney.2021, Gibney.2022}, in which the solution of the semidefinite program is computed at each SCF iteration with the splitting conic solver (SCS)~\cite{ocpb:16,scs} in the CVXPY Python program~\cite{diamond2016cvxpy}.  \textcolor{black}{We find that RDMFT generally has similar convergence to DFT with potentially more reliable convergence for strongly correlated molecules.}  The DFT calculations are performed in the Quantum Chemistry Package in Maple~\cite{rdmchem,maple} and PySCF~\cite{Sun.2018} while the RDMFT calculations are performed with a customized Python program that works with the Quantum Chemistry Package in Maple~\cite{rdmchem,maple} and PySCF~\cite{Sun.2018}.

\begin{figure}
    \centering
    \includegraphics[width=\linewidth]{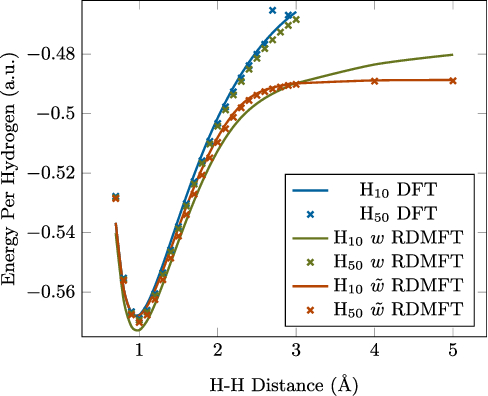}
    \caption{Energy per hydrogen atom as a function of the distance $R$ between equally spaced hydrogen atoms is shown for linear H$_{10}$ and H$_{50}$, using DFT with the SCAN functional as well as $w$ RDMFT and $\Tilde{w}$ RDMFT with the SCAN functional.} \label{fig:Hs}
\end{figure}

The energy per hydrogen atom as a function of the distance $R$ between equally spaced hydrogen atoms is presented for linear H$_{10}$ and H$_{50}$ in Fig.~\ref{fig:Hs}, using DFT with the SCAN functional as well as $w$ RDMFT, and $\Tilde{w}$ RDMFT with the SCAN functional.  For H$_{10}$, while both $w$ and ${\tilde w}$ RDMFTs improve upon the energy per hydrogen atom relative to DFT for $R>2$~\AA, only ${\tilde w}$ RDMFT exhibits the correct asymptotic behavior of a nearly flat potential energy curve for $R>3$~\AA.  Moreover, for the larger H$_{50}$, because $w$ is incorrectly decaying with system size, $w$ RDMFT shows minimal improvement over DFT, but ${\tilde w}$ RDMFT produces a curve that agrees with the one from H$_{10}$, exhibiting the expected convergence in the energy per hydrogen atom with respect to chain length in the dissociation limit.

\begin{figure}
    \centering
    \includegraphics[width=\linewidth]{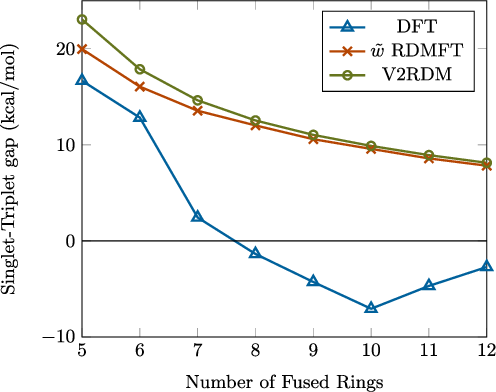}
    \caption{Adiabatic singlet-triplet gap---the energy difference between the lowest lying singlet and triplet states---for the $n$-acenes pentacene ($n=5$) through dodecacene ($n=12$), using $\Tilde{w}$ RDMFT with comparisons to DFT and V2RDM.  Both $\Tilde{w}$ RDMFT and DFT use the SCAN functional.} \label{fig:Acenes}
\end{figure}

\begin{figure*}[ht]
    \centering
    \includegraphics[width=0.98\linewidth]{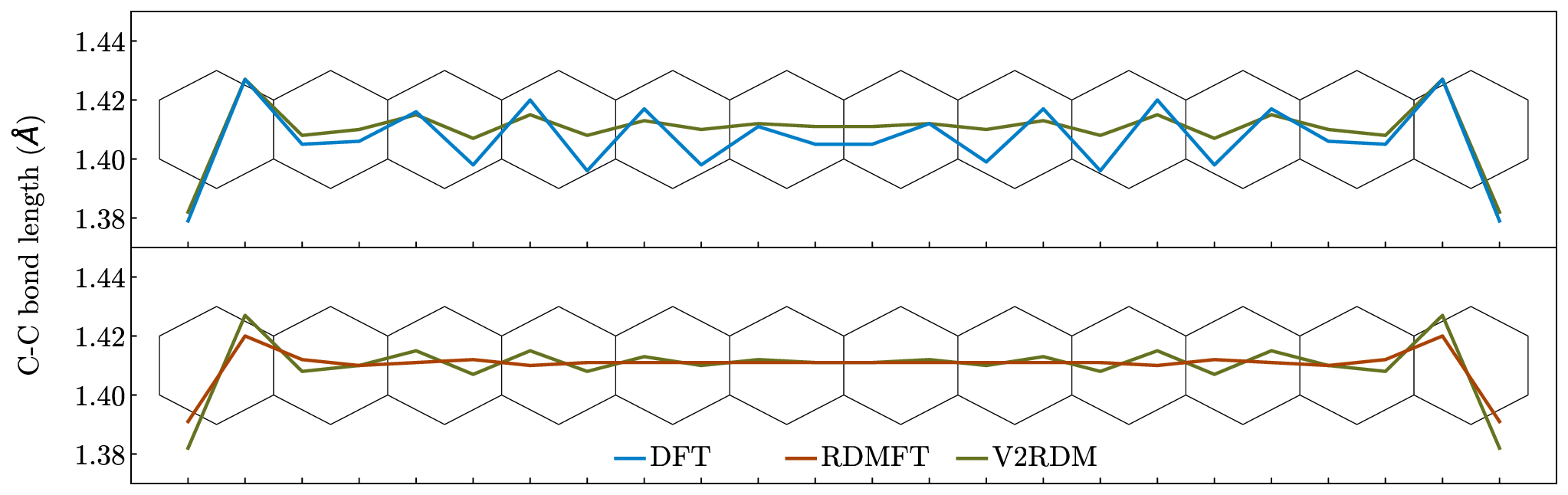}
    \caption{Edge carbon-carbon bond lengths from $\Tilde{w}$ RDMFT are shown in red with comparisons to those from DFT and V2RDM in blue and black, respectively.}
    \label{fig:Bonds}
\end{figure*}

Next, we consider the singlet-triplet gap---the energy difference between the lowest lying singlet and triplet states--- for the $n$-acenes pentacene ($n=5$) through dodecacene ($n=12$), displayed in Figure~\ref{fig:Acenes}.  The $n$-acenes have been shown to develop bi- and multi-radical character with increasing chain length~\cite{Hachmann.2007, Gidofalvi.2008, Hemmatiyan.2019, Schriber2018}. Due to the conjugated $\pi$ system increasing with the number of carbon atoms, acenes longer than tetracene are beyond the capabilities of traditional configuration-interaction-based methodologies such as the complete-active-space self-consistent-field (CASSCF) wave function method. Therefore, we compare our results to those from the \textcolor{black}{CASSCF} variational 2-RDM (V2RDM) method~\cite{Gidofalvi.2008, Mazziotti.2016co, Xie.20228s}, which replaces the calculation of the active-space wave function by configuration interaction by the calculation of the active-space 2-RDM by the V2RDM method with 2-positivity conditions~\cite{Mazziotti.2001g, Nakata.2001, Zhao.2004, Mazziotti.2004, Shenvi.2010, Verstichel.2011, Mazziotti.20115w, Mazziotti.2016co, Li.2021, Knight.2022, Xie.20228s, Mazziotti.2023}. The CASSCF V2RDM method---denoted in the following as just V2RDM---can treat the (50,50) active space required for dodecacene.  Relative to the results from V2RDM~\cite{Mullinax2019}, DFT underpredicts the gap for longer acenes with the triplet state becoming incorrectly lower in energy than the singlet state.  In contrast, the ${\tilde w}$ RDMFT closely matches the gap predicted by the V2RDM theory.  Extrapolating to infinite chain length $n$ using the function $a \exp{(-n/b)}+c$, we obtain singlet-triplet gaps of 7.23 and 7.77~kcal/mol from V2RDM theory and ${\tilde w}$ RDMFT, respectively.

Finally, we compute the equilibrium geometry of dodecacene in its ground singlet state with $\Tilde{w}$ RDMFT using analytical gradients.  In Fig.~\ref{fig:Bonds} the edge carbon-carbon bond lengths are displayed in red with comparisons to those from DFT and V2RDM, shown in blue and green, respectively.  The V2RDM method predicts edge carbon-carbon bond lengths that are fairly uniform in the interior of the acene; in contrast, DFT predicts an alternating bonding pattern with significant variations in the bond lengths.  The $\Tilde{w}$ RDMFT results, which exhibit fractional occupations, agree with those obtained from V2RDM in predicting equal bond lengths in the interior of the acene.  All edge C-C bond lengths from naphthalene to dodecacene for both the singlet and triplet states are available in the Supplemental Materials.

\textit{Conclusions:} DFT provides an $O(N^{3})$ electronic structure method that describes the electron correlation in many important molecules and materials, and yet it struggles to treat a specific type of correlation, known as static correlation, leading to errors in predicting charges, multiradicals, and reaction barriers. Here we enhance a recent generalization of DFT for static correlation to improve its treatment of this correlation for large molecules.  We renormalize a 1-RDM-based energy correction to DFT by using Cauchy-Schwarz inequalities of the electron-electron repulsion matrix to reweight the two-electron identity matrix, making the correction effective for large molecules.  The resulting RDMFT, while retaining the $O(N^{3})$ scaling of DFT, significantly improves upon DFT within statically correlated systems while reproducing DFT in systems lacking static, or strong, correlation.  We demonstrate its scalability with system size by considering a series of hydrogen and acene chains\textcolor{black}{, extending through} H$_{50}$ and dodecacene, respectively.  The theory offers new possibilities for calculating \textcolor{black}{the properties of} a broad range of strongly correlated molecular systems beyond the reach of multi-reference methodologies. \\

\begin{acknowledgments}

D.A.M. gratefully acknowledges the U.S. National Science Foundation Grant No. CHE-2155082.

\end{acknowledgments}

\bibliography{RDMFTR1.bib}

\end{document}